# Study on the fabrication of low-pass metal powder filters for use at cryogenic temperatures

**Sung Hoon Lee and Soon-Gul Lee**

*Department of Applied Physics, Graduate School, Korea University, Sejong City 339-700, Korea*

We fabricated compact low-pass stainless-steel powder filters for use in low-noise measurements at cryogenic temperatures and investigated their attenuation characteristics for different wire lengths, shapes, and preparation methods up to 20 GHz. We used nominally-30-μm-sized SUS 304L powder and mixed with Stycast 2850FT by Emerson and Cumming with catalyst 23LV. A 0.1 mm insulated copper wire was wound on preformed powder-mixture spools in the shape of a right-circular cylinder, a flattened elliptic cylinder and a toroid, and the coils were encapsulated in metal tubes or boxes filled with the powder mixture. All the fabricated powder filters showed a large attenuation at high frequencies with a cut-off frequency near 1 GHz. However, the toroidal filter showed prominent ripples corresponding to resonance modes in the 0.5-m-long coil wire. A filter with a 2:1 powder/epoxy mixture mass rate and a wire length of 1.53 m showed an attenuation of -93 dB at 4 GHz and the attenuation was linearly proportional to the wire length. As the powder-to-epoxy ratio increased, the high-frequency attenuation increased. An equally-spaced single-layer coil structure was found to be more efficient in attenuation than a double-layer coil. Geometry of the metal filter-case affected noise ripples with the least noise in a circular-tube case.

Email: sglee@korea.ac.kr

Fax: +82-44-860-1583

# I. INTRODUCTION

Low-temperature experiments, such as single-electron tunneling and superconducting qubits, are significantly affected by thermal and electrical noise. Accurate measurements in those experiments require elimination of such noises. The thermally-originated noise can be reduced by lowering the sample temperature and the external electromagnetic noise can be blocked by using a superconductor or μ-metal shield. However, the noise flowing in through the measurement circuits cannot be removed by those methods and thus should be eliminated by appropriate filters.

Various types of noise-eliminating filters have been suggested and fabricated. Several research groups have studied miniature thin-film filters [1, 2], distributed thin-film microwave filters [3], thermocoax filters [4, 5], and metal-powder filters [6-11]. A metal powder filter, which was suggested by Martinis et al. [6] for the first time, has shown attenuation of rf signals by skin-effect damping of the large surface area of powder. Performance of a metal-powder filter is known to be affected by composition (copper, bronze, brass, manganin, stainless steel, etc.), grain size, and the diameter and the length of wire. One can make metal-powder filters by using either metal-powder only or metal-powder/epoxy mixture. However, though all-powder filters have a better performance compared with its mixture counterpart, powder/epoxy mixture is used in cryogenic environments due to its relatively large thermal conductance which is important to avoid heating from rf dissipation. Recently, impedance-matched (50 Ω) powder filters [9] and powder-on-printed-circuit-board (PCB) filters [11] have also been studied. Metal-powder filters are advantageous in its easy fabrication, but also frequently show unwanted resonance ripples.

We studied fabrication of low-pass stainless-steel powder filters for use at cryogenic temperatures and investigated the effects of the powder/epoxy mixture ratio, wire length, and the filter shape on the performance of the filter. We measured attenuation properties up to 20 GHz by using a vector network analyzer (VNA) down to 4.2 K. We also investigated various preparation methods to minimize the unwanted resonance ripples.

## II. SAMPLE PREPARATION

Generally, a cryogenic low-pass metal-powder filter is made by (1) preparing a copper-wire coil wound on a rod made from a mixture of metal powder and low-temperature epoxy, (2) inserting the coil in a metal case which is afterwards filled with the same powder/epoxy mixture without a gap or a void inside the case, and finally (3) connecting the wires to rf connectors mounted on the case. As for the low-temperature epoxy, Stycast 1266 or 2850FT are commonly used. Stycast 1266 has a low viscosity, which is advantageous for increasing the powder ratio to enhance the attenuation effect. On the other hand, 2850FT has a relatively large thermal conductivity at low temperatures, which is important to avoid thermal pileup inside the filter due to rf power. For these reasons, a mixture of 1266 and 2850FT has also been used [9]. Attenuation characteristics of a powder filter is affected mostly by the resistivity of the metal composing the powder. Powders of various metals, such as copper, stainless steel (SUS), bronze, and manganin, have been adopted to make filters. Among those, SUS powder produced most efficient attenuation, presumably due to a relatively large surface resistance.

We prepared coil-spool rods of the shape of a right-circular cylinder, a flattened cylinder, and a toroid. The spool rods were made from SUS-powder/epoxy mixtures with different mixture ratios. We used SUS 304L powder of nominally 30 $\mu$m in diameter and Stycast 2850FT supplied by Emerson and Cumming with catalyst 23LV. An insulated copper wire of 0.1 mm in diameter was wound on preformed powder-mixture spools and encapsulated in metal tubes or boxes filled with the powder/epoxy mixture. Finally, ends of the coil wire were connected to SMA connectors mounted on the filter case. We also prepared filters with several different lengths of the coil wire. The filter cases were circular tubes made of copper and square boxes made of stainless steel.

Figure 1 shows images of the surface (top) and the cross-section (bottom) of right-cylindrical spool rods made from a powder/epoxy mixture with 2:1 mass ratio. We discarded rods containing voids (left) and used those without voids (right) in our experiments. Formation of voids was avoided by pumping gas bubbles from the mixture for a sufficient period of time. We used polymer straws for the cylindrical and elliptic spool rods and SUS molds for toroidal spools. Figure 2 shows images of coils wound on the spool of the shape of (a) a right-circular cylinder, (b) a flattened elliptic cylinder, and (c) a toroid. Each coil was made of a 0.5 m-long insulated copper wire of 0.1 mm in diameter and wound uniformly on spools to minimize reduction of the attenuation effect due to capacitive coupling [8]. We inserted the coils in the filter cases, connected ends of the coil wires to SMA connectors, and filled the cases with the same SUS-powder/epoxy mixture without a gap. Voids or gaps must have been removed to avoid unwanted resonances. Figure 3 shows completed filters. The circular-cylindrical coil filter was encased in either a circular tube or a square box, and the other types of filter were encased in a square box.

## III. EXPERIMENTL RESULTS AND DISCUSSION

Attenuation properties of metal-powder filters were measured by using a spectrum analyzer (HP 8596E) and a vector network analyzer (Agilent E5071C) at room temperature, 77 K, and 4.2K. As shown in Fig. 4, all of the three filter types exhibited increase in attenuation with increasing frequency. The magnitude of attenuation at 1 GHz is -20 dB at room temperature [Fig. 4(a)] and -15 dB at 77 K [Fig. 4(b)]. All the data show resonance ripples with a fundamental frequency of 0.2 – 0.22 GHz. Among the three filter types, the toroidal filter shows most prominent resonance peaks at multiples of 0.22 GHz, which is believed to be due to an electromagnetically-linked end structure of the toroidal coil. From the resonance condition in the wire, the effective speed of rf signals in the filter is calculated to be $0.67c$ for the circular solenoid, $0.70c$, for the flattened-elliptic solenoid and $0.73c$ for the toroid filter, where $c$ is the speed of light.

Figure 5 shows attenuations at room temperature for filters with different lengths of the copper coil-wire: 0.5 m, 1.0 m, and 1.7 m. All the filters had square-box cases. The figure shows that the attenuation characteristics increases with increasing wire-length with a magnitude of attenuation at 1GHz of -17 dB for 0.5 m, -32 dB for 1.0 m, and -55 dB for 1.7 m. These results are in agreement with the studies of Fukushima et al. [7] and Lukashenko et al. [10]. However, for a given length of coil wire, the configuration of the coil also significantly affected the filter performance.

Fig. 6 shows the performances of right-circular cylindrical filters with different coil configurations: (a) a single-layer coil of 1.533 m in length and 6.39 turns/mm in turn density and (b) a double layer of 2.985 m in length and 12.22 turns/mm in turn density. The filters had circular-tube cases. To make the double-layer coil, we wound the first-layer coil on a 3-mm mixture-rod, then coated the 3-mm coil with the same powder/epoxy mixture to make a 5-mm rod, and finally wound the second layer coil on it in the same direction. As shown in the figure, though the single-layer coil has a relatively shorter wire length, its attenuation property is much better than the double layer-coil filter. The magnitudes of attenuation decreased at 4.2 K compared with those at room temperature for both filter structures, but the overall tendency of the coil-configuration dependence is the same. These results indicate that the configuration of a filter coil is an important factor determining the performance of a metal-powder filter.

Geometrical structure of a filter case affected noise properties of a powder filter. Depending on the symmetry of the space around the filter coil, complicated multiple resonances can be generated and appears as noise ripples in attenuation curves. The data of Figs. 5 and 6 were taken from right-circular solenoid filters that were cased in square boxes and circular tubes, respectively. For all the filters in the figures, coils were prepared in the same way and placed in the middle of the cases. As shown in the figures, the ripples observed in the attenuation curves of square-box-case filters (Fig. 5) are more prominent than those of circular-tube-case filters (Fig. 6), for which the ripples are almost not observable above 0.4 GHz. Similar case-geometry-dependence of noise ripples can be inferred from studies by different research groups: ripples observed in a square-box filter by Mueller et al. [11] were larger than those observed in a circular-tube filter by Lukashenko et al. [10].

In Fig. 6, the attenuation of a circular-coil filter decreases with increasing frequency from -32 dB at 1GB to -110 dB at 4.3 GHz, and then decreases again at higher frequencies. Such decrease in attenuation at higher frequencies is ascribed to the low density of SUS powder in the mixture due to high viscosity of Stycast 2850FT and could actually be improved by increasing the powder density in the mixture. For comparison, we prepared a filter made of SUS powder only and measured its attenuation characteristics. As shown in Fig. 7, the powder-only filter has a much better filtering performance, reaching the noise floor of -110 GHz above 2 GHz and the humps above 4 GHz, observed in the powder/epoxy mixture filter, disappear. At 1 GHz, the powder-only filter has an attenuation of -70 dB, which is better than its mixture counterpart by more than twice even if the wire length is half of the mixture filter. This result implies that the density of the metal powder in the mixture plays an important role in enhancing attenuation of the filter.

## IV. CONCLUSION

In conclusion, we have studied fabrication of compact low-pass stainless-steel powder filters for use at cryogenic temperatures and investigated their attenuation characteristics for different wire lengths, filter shapes and preparation methods. All the fabricated powder filters showed a large attenuation at high frequencies with a cut-off frequency near 1 GHz and the attenuation was linearly proportional to the wire length. Among the coil types we have studied, the right-circular cylindrical solenoid showed the best performance with the least noise ripples. The configuration of the filter coil was also important in determining the performance of a metal-powder filter; an equally-spaced single-layer coil structure was more efficient than a double-layer-coil filter. Geometry of the metal filter-case affected noise ripples with the least noise in the circular-tube-case filter. We also observed that the high-frequency attenuation increased as the powder-to-epoxy ratio increased.

## ACKNOWLEDGEMENT

This research was supported by a Korea University Grant.

**Figure Captions.**

Figure 1. Images of coil-spool rods made from a mixture of SUS 304L powder and Stycast 2850FT with a 2:1 mass ratio. Rods containing voids (left) were discarded and those without voids (right) were used in our experiments.

Figure 2. Images of coils wound on the spool of the shape of (a) a right-circular cylinder, (b) a flattened cylinder, and (c) a toroid. Each coil was made of a 0.5 m-long insulated copper wire of 0.1 mm in diameter.

Figure 3. Completed filters made of (a) a flat solenoid and a toroid and (b) a right-circular cylindrical solenoid. The square-box case in (a) is made of stainless steel and the circular-tube case in (b) is made of copper.

Figure 4. Frequency-dependent attenuation characteristics of SUS powder filters measured at (a) 300 K and (b) 77 K. Nominal size of the SUS powder was 30 μm in diameter.

Figure 5. Attenuation characteristics of the powder filters of the shape of a right-circular cylindrical solenoid for different wire lengths: (1) 0.5 m, (2) 1.0 m, and (3) 1.7 m. The filters had square-box cases made of stainless-steel.

Figure 6. Attenuation characteristics of the right-circular cylindrical filters with different number of layers and turn densities: single-layer with 6.39 turns/mm (total length = 1.53 m) and double layer with 12.22 turns/mm (total length = 2.98 m). The filters had circular-tube cases made of copper.

Figure 7. Attenuation characteristics of the right-circular-cylindrical-coil filters with different powder/epoxy mass ratios: 2:1 mixture and all SUS powder.

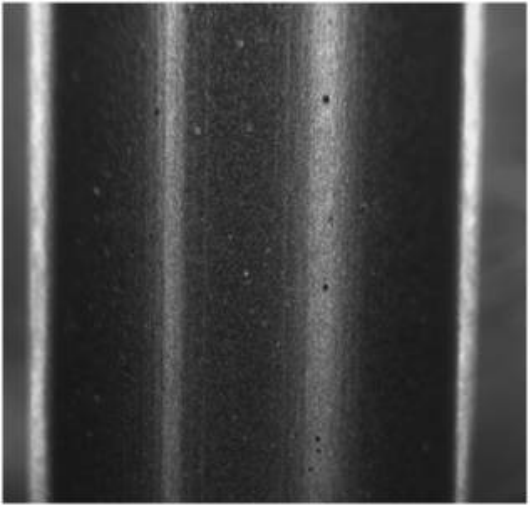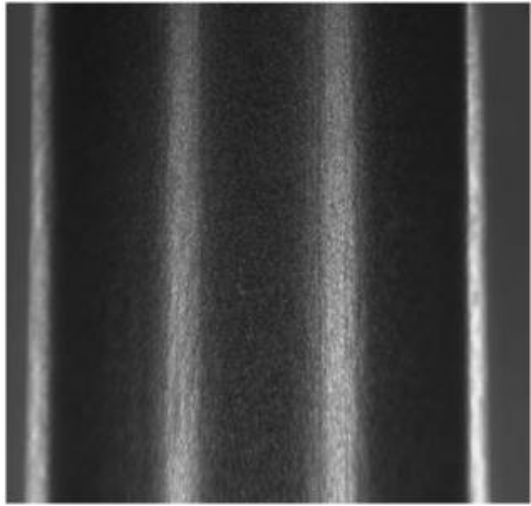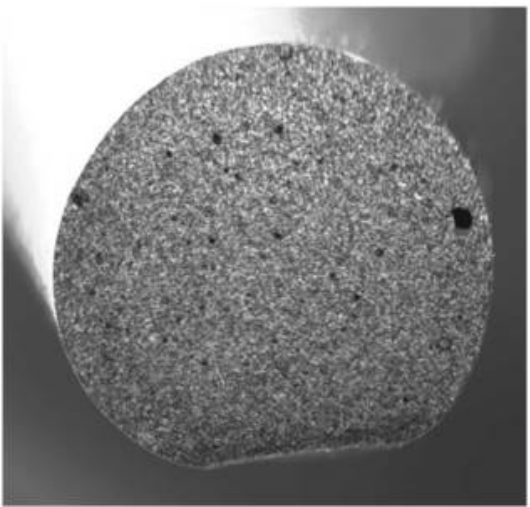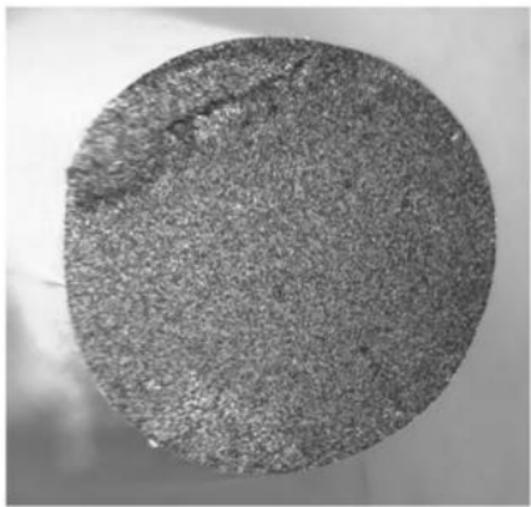

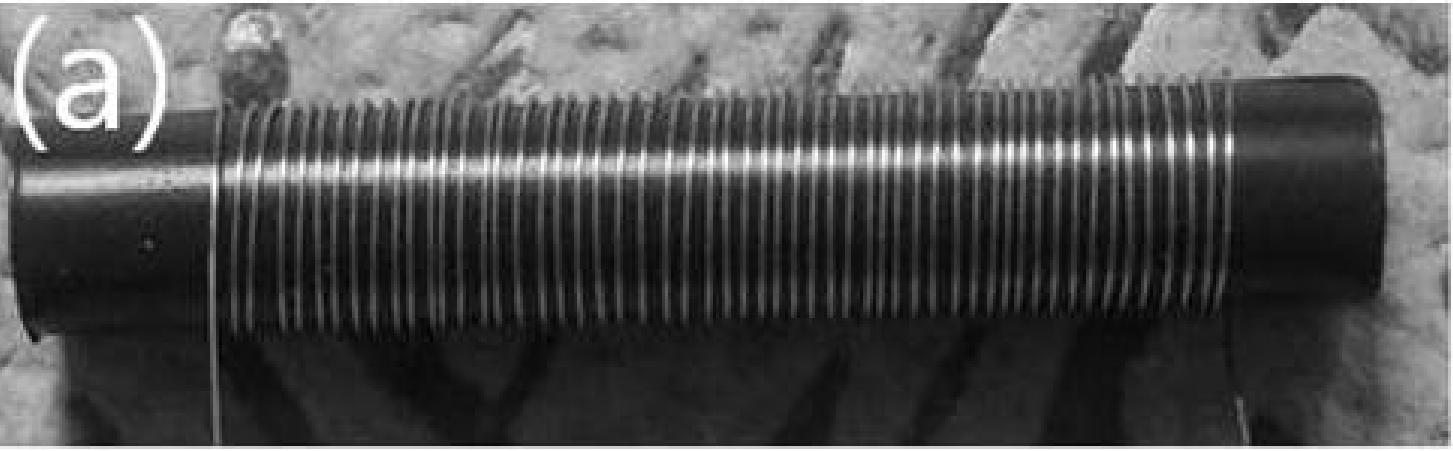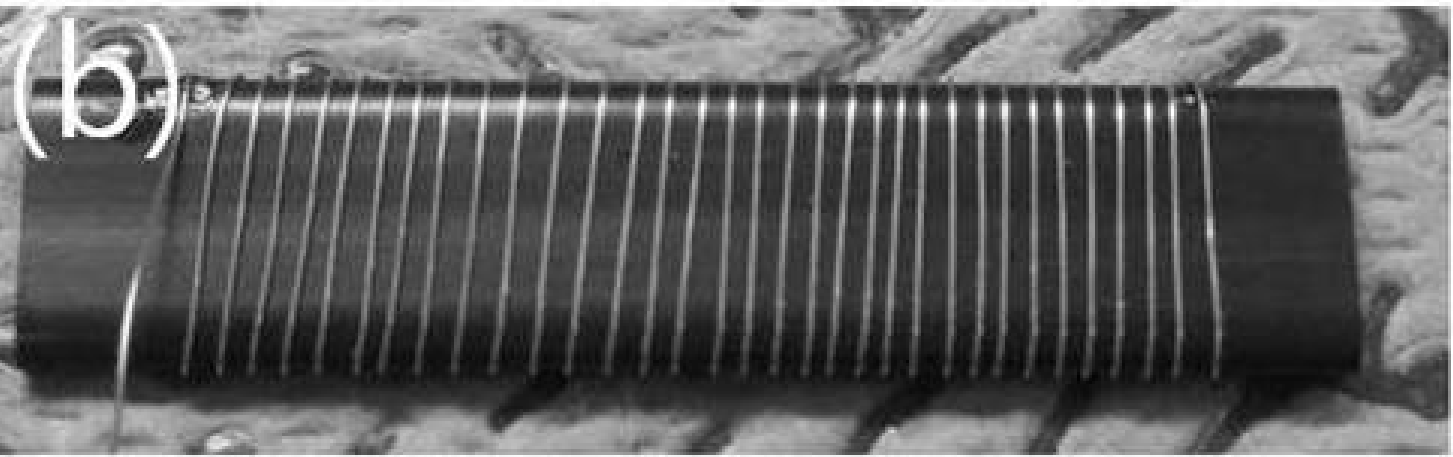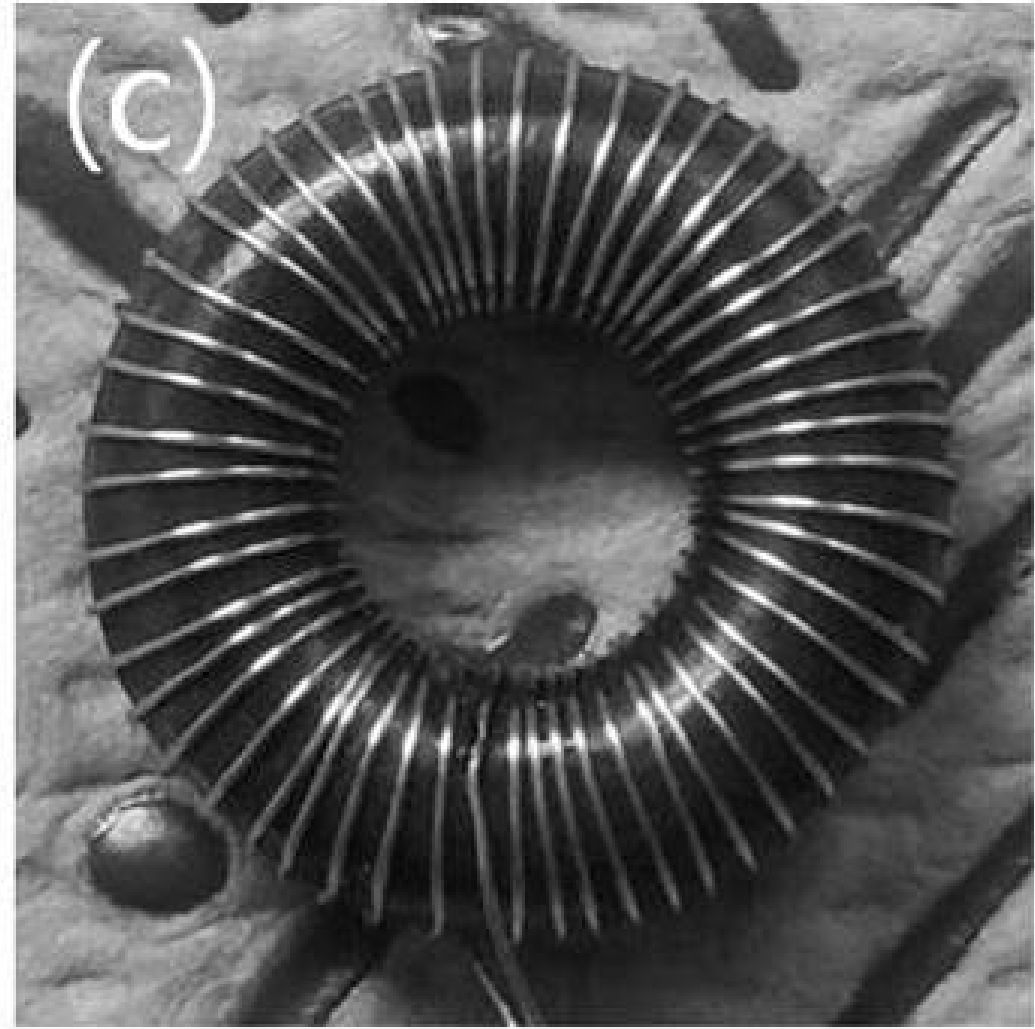

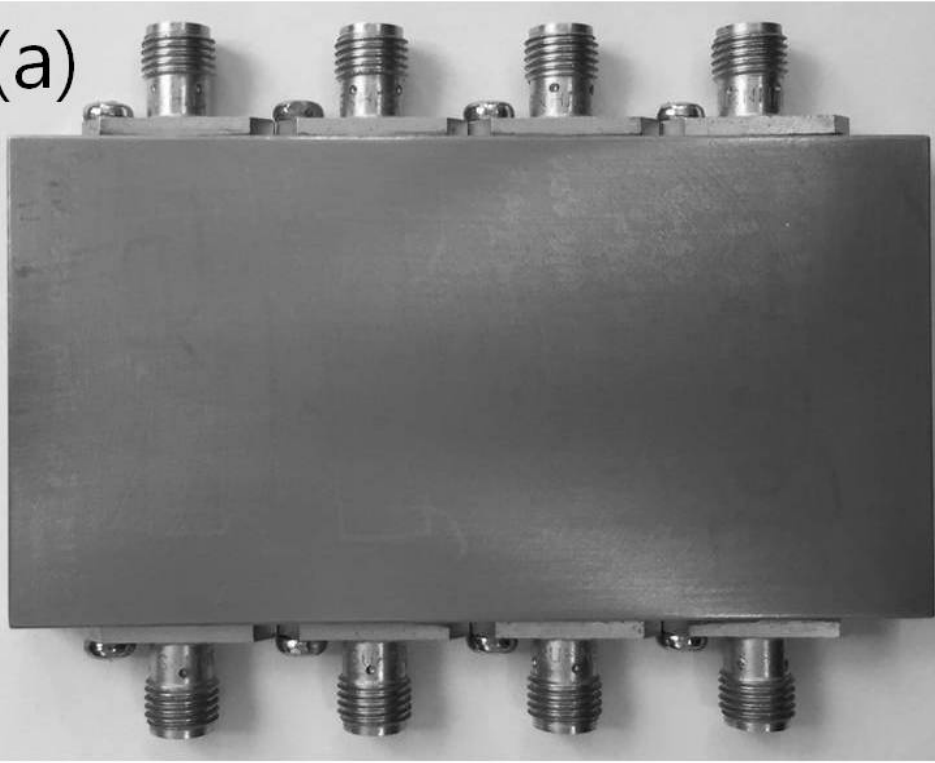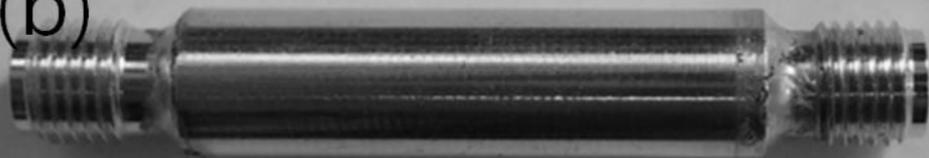

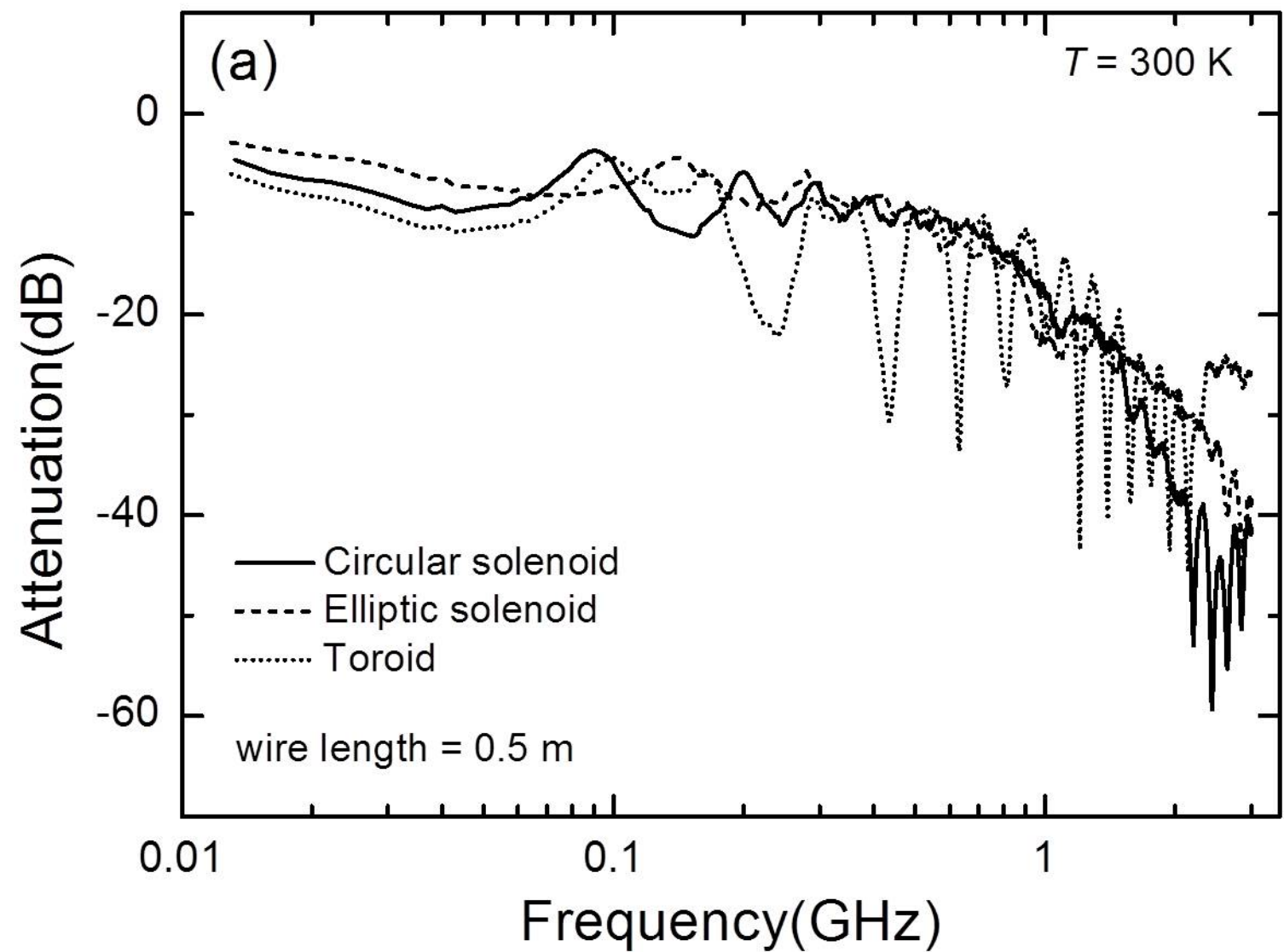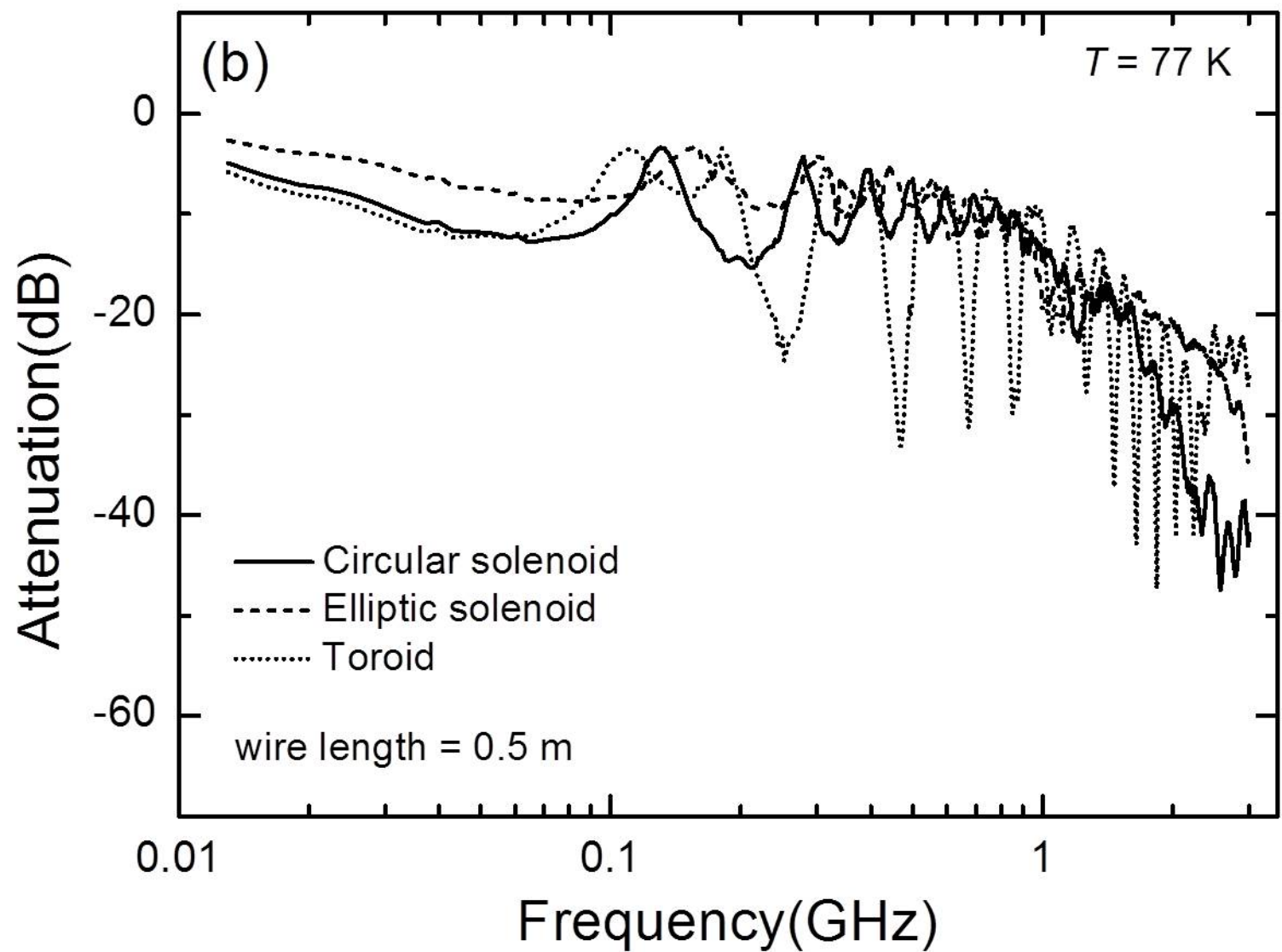

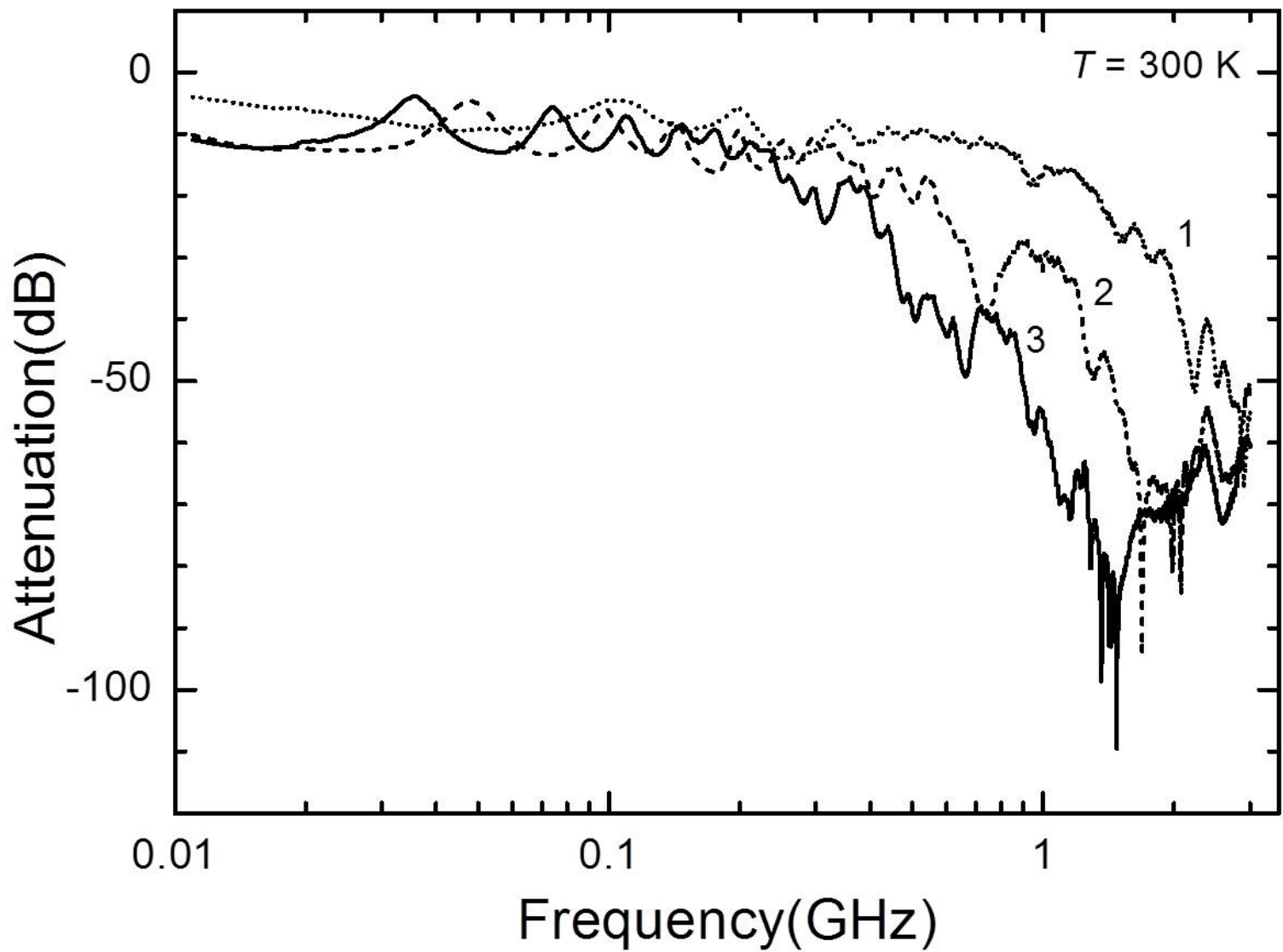

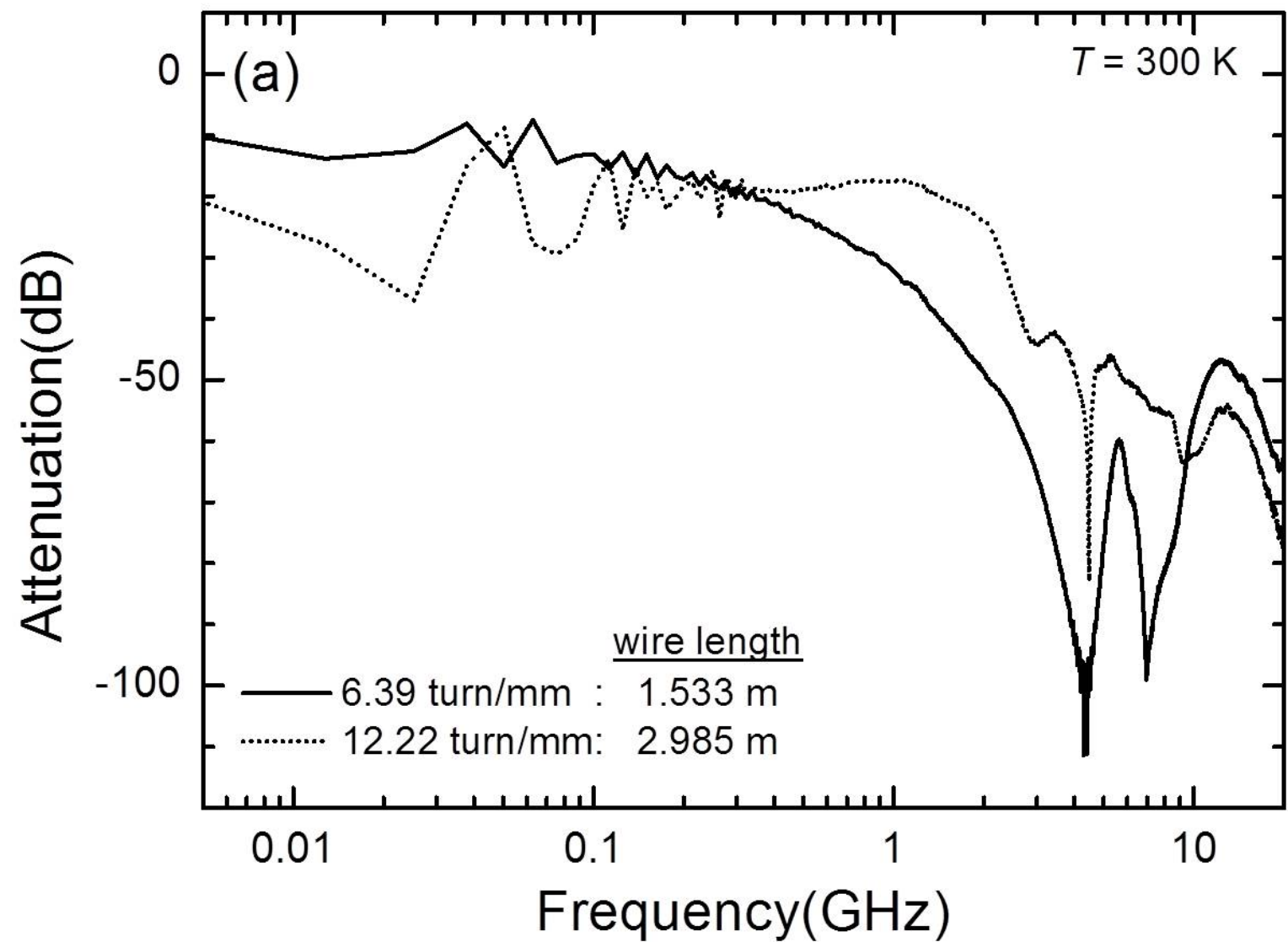 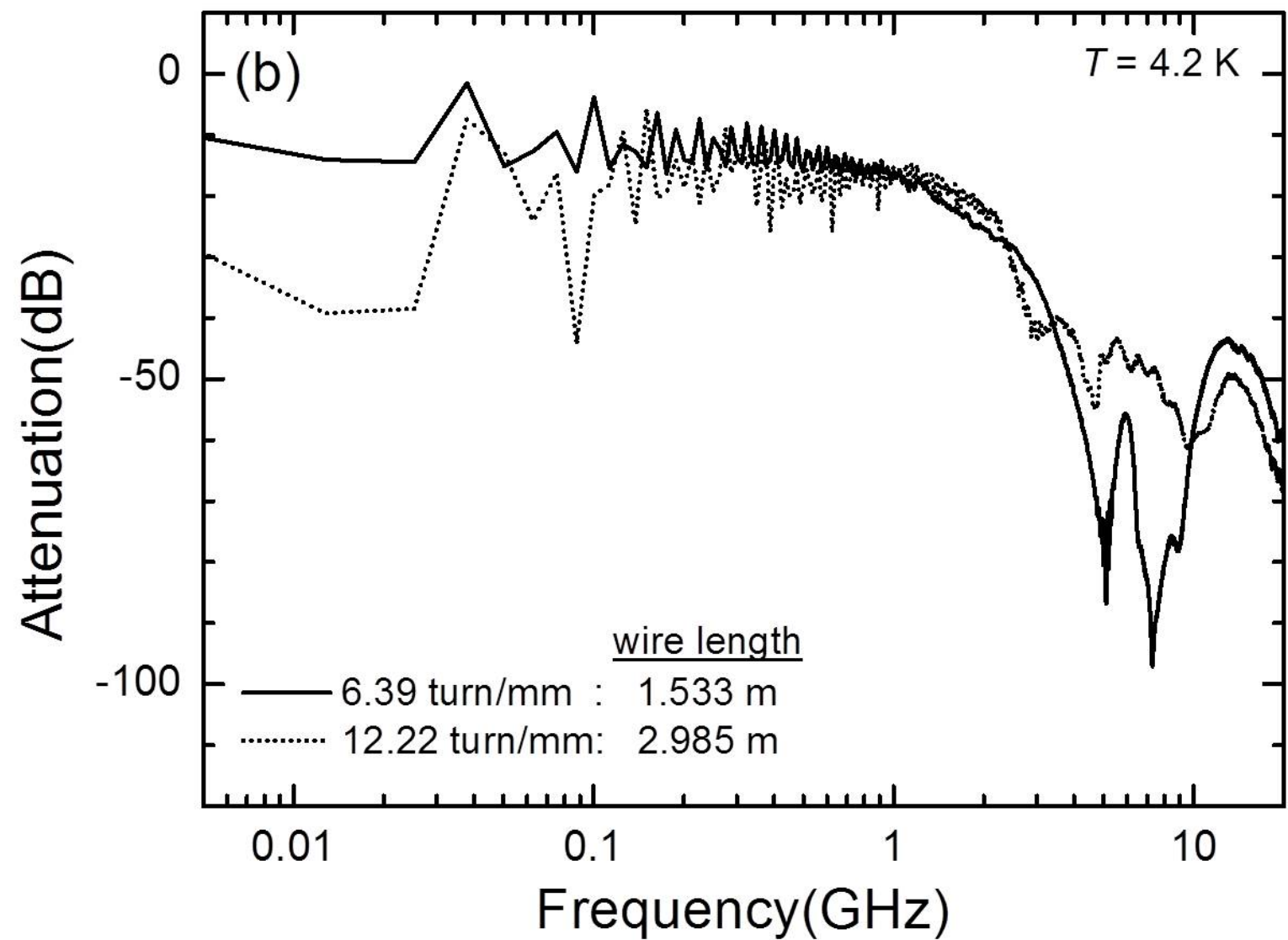

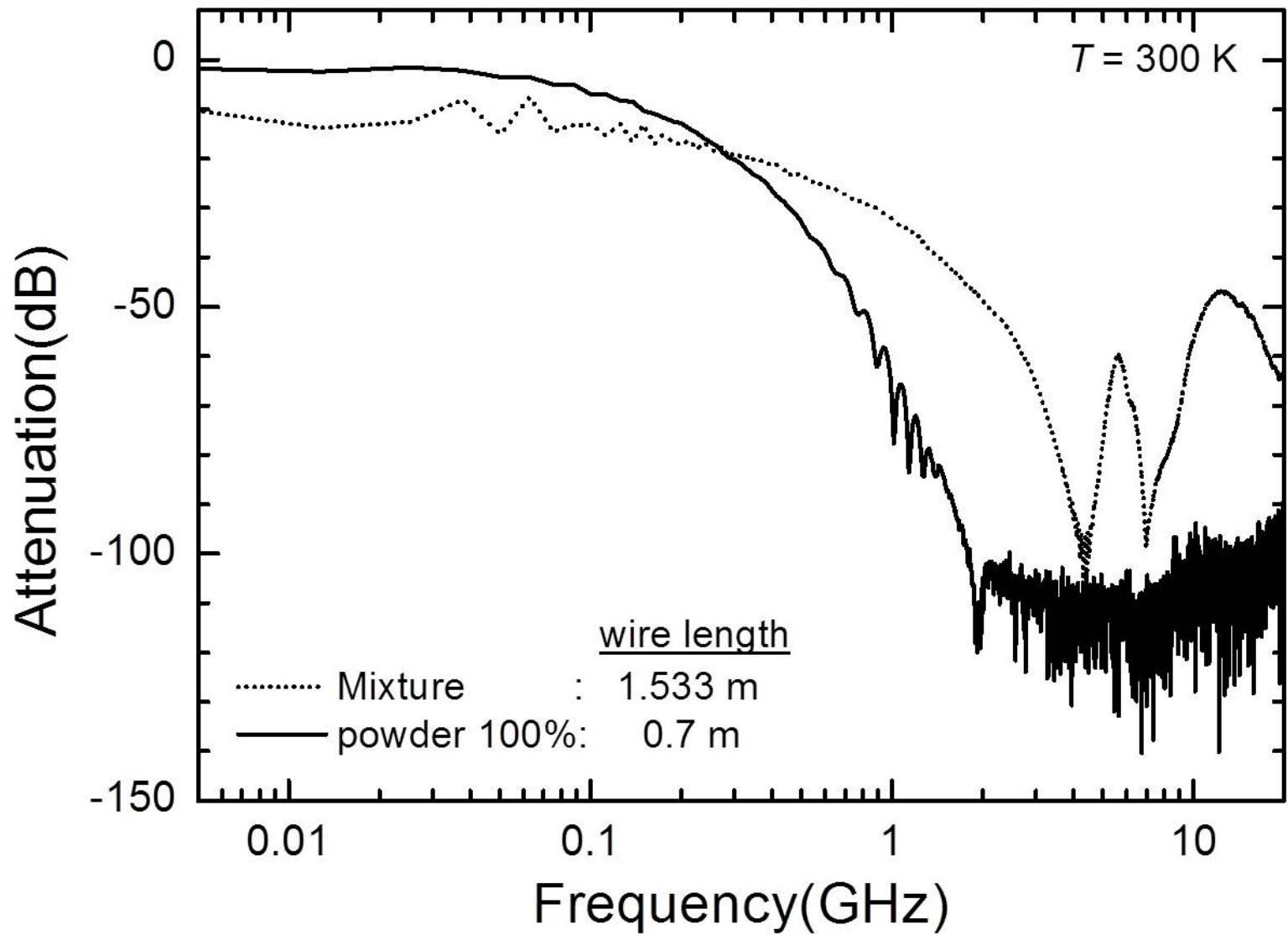